# Layered structure and leveled function of a human brain

Shengyong Xu, Jingjing Xu, and Rujun Dai

Department of Electronics, School of Electronics Engineering and Computer Science, Peking University, Beijing, 100871, P. R. China

**Introduction**

If stretched out to a flat plain, the cortex of a brain in an adult human may have a surface area of around 2,000 square centimeters. It contains 1-2 billion neurons in six distinguishable layers, adding up to the 2-3 mm thick "grey matter". For the rest of the brain, its majority volume is made of "white matter" of neural fibers (axons and dendrites) and other cells, e.g., glial cells, which provide brain homeostasis and may help neurons to live and work well [1, 2]. On the cortex, researchers have recognized over two hundred local areas that specially assign for different functions and may excite under different statuses [3, 4]. Recently, it is found that intensively arranged parallel axons form effective isolation walls for separate motion and diffusion modes of intracellular fluidics in a brain [5]. Yet to date the working mechanisms of the cortex and human brain remains not fully clear.

**Localization effect in layered cortex**

Fast reactions of a single neuron cell are either being exciting for ~ 1 ms as measured, or being idle. Chemical processes in a neuron are much slower. Therefore, to store complicated data in brain functions it needs a large group of neuron cells, instead of only one. We presented a memory model recently, that "2D codes" made of geometric patterns of a group neurosomes which can be excited at the same time, can be used as carriers for memory data [6]. According to this model, to maintain a 2D code for a while for sensing, analyzing and memorizing, it requires a vertical coupling mechanism between two neighboring layers of neurosomes. A strongly connected 2D pattern of neurosome is exciting for around 1 ms, then by vertical dendrites and synapses, the exactly same pattern in the next layer is triggered exciting after a short delay. Next, the original pattern in the first layer is triggered exciting again with the same vertical dendrite paths. This "echoing" process may last for seconds or longer, appearing as "temporary memory", until the external stimuli stops, or the active learning process is completed. Repeating this process, a temporary memory, i.e., a 2D code, can be consolidated and gradually turn into long-term memory, where synapses connecting the neurosomes in this 2D pattern all turn into strongly connected electrical synapses.

If a neurosome-based 2D code represents complex data, e.g., part of an image, it may consist of hundreds and thousands of neurons. Physically it seems impossible to transfer horizontally such a complex pattern from one location to the other on the cortex, because it needs a huge number of horizontal axons that beyond the feasible structure of a layered thin cortex in brain. This inference is consistent with anatomy studies on the brain. It is an important fact. It implies that data proceeding activities for different functions in a brain are most likely localized, limited to certain specified cortex region. It also implies that the detailed data of long-term memories and temporary memories for individual sensory and analysis, e.g., for an image or a piece of music, are stored in the same region of the cortex, but may be in different layers of the region. [6].

**Leveled zones for different functions**

A brain is a parallel processer. Sensing and analyzing images, sounds, smells, touches, temperatures, etc., as well as reacting with voices and body motion, may all proceed at the same time in parallel. In the cortex, individual local regions assigned for these parallel different functions separate with each other by a distance of a few centimeters to tens of centimeters. As just discussed, to transfer complex information occurred at one location of the cortex horizontally to another region is physically not feasible. Yet parallel processing is commonly working well for most people. It raises a question: How does the owner of this brain, a gentleman for instance, remember the events that he is listening to Leonardo Cohen, talking to his wife next to him, watching passing cars on the road and at the same time when he drives a Cadillac?

Parallel processing, analysis of multi-channel information, events, etc., require a region in the brain that receives all the information at the same time. This region should not be somewhere in the cortex. In this region, neurons mix up the multi-channel data and then make complicated reaction such as reasoning and concluding, creating a happy mood, starting to run. For connecting to multi-region of the cortex, it needs a large number of long and parallel axons which take a huge volume of the "white matter". In particular, to save the total length of "axon cables", this region is likely located in the center of the brain. Cingukate gyrus, hippocampus, etc., seems good candidates for such functions of "overall analysis".

In such a multi-functional region, the neurons are supposed to react (i.e., excite) with "secondhand information", if what occurs at different region of cortex is defined as the "firsthand information". Compared to the cortex, this region has a much smaller surface area, thus much less neurosomes. Following the "2D code" model, its storage capacity for "events" and other data is also much limited. As a reasonable assumption, this region stores only important data for surviving and for intelligence development. The details of an individual event may still be stored at corresponding region of the cortex, where once triggered from signals sent by central regions of the brain through axons, these detailed 2D codes for the event details can be retrieved.

Next question: How to judge the importance of an event and then make decision either keeping it as data in long-term memory, or deleting it? Similarly, such a judgement could not be made with neurons in the same layer. It needs another region of the brain, either in different layers of the hippocampus, for instance, or a smaller organ in the brain other than the hippocampus. In either way, this "higher level" neuron clusters need to access to information about outputs of "event analysis", instead of "detailed sensory" and "events" themselves.

Thus, we clearly see that the brain is a highly "layered structure" in the cortex, and probably in other parts like the hippocampus. Meanwhile, it has "leveled functions", where different zones react to the firsthand information (e.g., image and sound), secondhand information (e.g., events), and so on, respectively.

Comparison of the outputs from events, i.e., analyzing the time sequence of events and consequences of reactions, moods, body reactions, etc., is supposed to occur at the third level, in a smaller area of the limbic system. There might exist a higher level, the fourth level, which takes charge of analyzing the output of the third level, sending feedbacks to the third and lower levels. The outputs of the fourth level function may make up "consciousness", those differ a human from animals, and one person from the other. Probably, the mindsets of "thinking", "reasoning", as well as "ego", might be treated and stored at this "highest level" of neurons, a "core of the brain", which is smaller than the moderate hippocampus. The "core" might be amygdala, in viewing the location, size and connection to the rest part of the brain.

**Basic reaction modes: comparison and imitation**

To make a decision, a brain relies on the reaction of its neurons. Neurons have only limited reaction modes, such as exciting, idle, or releasing certain molecules at synapses. If 2D codes of neurosome are indeed the information carriers, the judgement output of neurons in certain layer of the cortex may result from the comparison of two 2D codes, e.g., one received from sensor such as an image just seen, and the other retrieved from memory [6]. By overlapping these two codes on a third layer of neurons, we assume that those matching points could enhance excitation, and nonmatching points lead to no reaction. The pattern of exciting neurosomes on this comparison layer is a measure of the comparison result. The more similarity of the two 2D codes under comparison, the higher the total excitation intensity is.

Repeating external stimuli leads to enhanced memory. Comparison of new sensing information or experience with the existing memory leads to decisions: A positive comparison result means a high level of similarity and it may favor one kind of body reaction, such as a comfort and happy mood when recognizing a familiar face; a negative comparison result may lead to other kinds of body reaction, such as ignoring.

Just like the enhancement effect of echoing mechanism (a kind of feedback) in memory, feedback mechanism is crucial in comparison analysis, without which it could not be explained why the processes like analysis, judgement and decision sometimes need a longer time than 1 ms exciting time in neuron. The feedback mechanism involves an important and complex issue of "matching of time scales in brain function" that needs further discussion.

Numerous facts show that, many animals have basic memory functions and comparison reactions for making judgments. Both human and animals have the capability of imitation in their brain processes and body behaviors. In many species, individuals with bigger and better features (such as stronger, faster, more beautiful) are admired and imitated. The underlying mechanism for this universal phenomenon might have played crucial roles in evolution of life on this planet. Yet among all creatures, only human developed complicated education system to enhance the imitation processes to the extreme levels, which results in the accumulation of human intelligence that beyond the imagination of any other animals.

**Language, the key role in human intelligence**

Many birds are good singers. Some may learn the voice of human. Some animals, such as whales, dolphins, elephants, apes, can communicate each other with a number of simple sounds. Only human developed complicated languages. One well-developed human language has thousands of words, and combinations of these words can describe millions of daily subjects, abstract concepts, and events. A single sound could represent a very complicated context, which is precisely defined and can be accepted and remembered by a large group of people through education and communication. To memorize a single sound, it needs much less neurosomes in the 2D code. This remarkably saves the memory space for individual brain, so that one brain can store a large volume of the results of civilization, thus offer a huge pool of elements for further analysis, comparison, imitation, and logic thinking, which has been greatly promoting the development of human intelligence. Just as Sapir-Whorf hypothesis states: high level thinking process is based on language [7-9]. With the help of language, human established commonly acceptable abstract concepts, imparted the past survival skills from one generation to the next, and communicated religions to strongly connect thousands and millions of people.

**Symbols and tools, extension of a drain**

In a time span of one million years, human invents a huge number of symbols and tools. Symbols, such as written characters, arts and music, took a major memory job for the entire data of human intelligence. A library with millions of books is a typical collection of symbols invented by human, showing the data of life experiences of millions individuals in hundreds of generations. Anytime by education or learning process one human brain can check and compare its own memories and life experiences with these the external symbols. The "external brain memory" have connected the brain of an individual human to the brains of millions of outstanding humans ever lived. Knowledge of all disciplines, such as mathematics, physics, social rules, laws and religions, are also symbols, invented piece by piece by individual brains in the whole history. Numerous tools and techniques associated with the knowledge have incredibly extended the capability of both brain and body of a human being towards an unlimited level. The development of computers and internet obviously accelerates this process.

By sharp contrast, the functions of a single brain are very much limited. It is the summary of the efforts of billions people who ever lived and still living on this planet that established the entire building of civilization. Languages, music, arts, tools, social rules, sciences, religions, and so on. For an individual brain, it can only make rational judgements by comparing his or her ongoing experiences with the background of the civilization. It may make creative contributions, but not much. It has limited memory capacity; occupying too much room for memory in an individual brain, for instance, by years of aggressive education, may degrade its function capacity.

**Summary and impact**

The anatomically layered structure of a human brain results in leveled functions. The first-level function is performed at the cortex, neuron clusters in spinal cord, etc. Neurons here are in charge of sensory, body motion, as well as memory of selected details of these functions, most likely in 2D codes of neurosome patterns. The second level collects analytical outputs from neurons at different local regions of the first-level structure via long axon arrays, thus data of "events" are created and stored in this region. Comparison of the outputs from events, i.e., analyzing the time sequence of events and consequences of reactions, etc., is supposed to occur at the third level, in a smaller area of the limbic system. There might exist a higher level, the fourth level, which takes charge of analyzing the output of the third level, sending feedbacks to the third and lower levels. The outputs of the fourth level function may probably make up "consciousness", "ego", etc. In all these levels of different functions, comparison, feedback and imitation are the universal and crucial mechanisms. Languages, symbols and tools play key roles in the development of human brain and entire civilization.

## Acknowledgements

We thank Dr. J. Tang for valuable discussions. This work is financially supported by The Seed Joint Project of MD & EECS 2016, Peking University. (Email: xusy@pku.edu.cn).